\documentclass[conference]{IEEEtran}
\IEEEoverridecommandlockouts

\ifCLASSINFOpdf
  \usepackage{graphicx}

\else
 
\fi

\usepackage{amsmath}
\usepackage{algorithm,algorithmic}
\usepackage{amssymb}
\usepackage{cite}

\newtheorem{proposition}{\bf Proposition}

\newtheorem{definition}{\bf Definition}

\begin{document}
\title{Collaborative Artificial Intelligence (AI) for User-Cell Association in Ultra-Dense Cellular Systems}


\author{\IEEEauthorblockN{Kenza Hamidouche\IEEEauthorrefmark{1},
Ali Taleb Zadeh Kasgari\IEEEauthorrefmark{2},
Walid Saad\IEEEauthorrefmark{2}, 
Mehdi Bennis\IEEEauthorrefmark{3} and
M\'erouane Debbah\IEEEauthorrefmark{1}\IEEEauthorrefmark{4}}
\IEEEauthorblockA{\IEEEauthorrefmark{1} LSS, CentraleSupelec,
Universit\'e Paris-Saclay, Gif-sur-Yvette, France Email: kenza.hamidouche@centralesupelec.fr.}
\IEEEauthorblockA{\IEEEauthorrefmark{2} Wireless@VT, Bradley Department of Electrical and Computer Engineering, Virginia Tech, Blacksburg, VA\\
Emails:\{alitk,walids\}@vt.edu.}
\IEEEauthorblockA{\IEEEauthorrefmark{3} CWC - Centre for Wireless Communications, Oulu, Finland, Email: bennis@ee.oulu.fi}
\IEEEauthorblockA{\IEEEauthorrefmark{4} Mathematical and Algorithmic Sciences Lab, Huawei France R\&D, Paris, France Email:  merouane.debbah@huawei.com.}
\thanks{This research was supported by the U.S. National Science Foundation under Grants CNS-1460316 and IIS-1633363.}}


\maketitle

\begin{abstract}
In this paper, the problem of cell association between small base stations (SBSs) and users in dense wireless networks is studied using artificial intelligence (AI) techniques. The problem is formulated as a mean-field game in which the users' goal is to maximize their data rate by exploiting local data and the data available at neighboring users via an imitation process. Such a collaborative learning process prevents the users from exchanging their data directly via the cellular network's limited backhaul links and, thus, allows them to improve their cell association policy collaboratively with minimum computing. To solve this problem, a neural Q-learning learning algorithm is proposed that enables the users to predict their reward function using a neural network whose input is the SBSs selected by neighboring users and the local data of the considered user. Simulation results show that the proposed imitation-based mechanism for cell  association converges faster to the optimal solution, compared with conventional cell association mechanisms without imitation.
\end{abstract}


% \fi
%Hamilton-Jacobi-Bellman equation:

\IEEEpeerreviewmaketitle

\section{Introduction}

The emergence of the Internet of Things (IoT) has given rise to a significant amount of data, collected from sensors, user devices, and base stations, that must be processed by next-generation wireless systems \cite{Park_learning,Yaqoob_IoT,Luong_data,moz,abu}. Relying on traditional cloud-centric approaches for big data analytics may no longer be suitable for dense cellular systems that encompass both IoT devices and conventional mobile phones. Instead, it has become imperative to leverage the distributed storage and computing power available in the network infrastructure and devices (e.g., smartphones, computers, tablets and base stations) so as to process the data. The data that is gathered at the devices and base stations (BSs) is primarily related to the network operations and includes the number of connections from a given device to a BS, the type of requested data, the traffic load at specific time periods, the location of the users, and the channel state information, among others. Clearly, such type of data is private in nature and users or BSs that are owned by different network operators would be reculant to share their own collected data. 

Recently, machine learning-based artificial intelligence (AI) techniques \cite{Chen_AI,Kasgari2017Asilomar} have emerged as promising tools that allow a cellular network to leverage the aforementioned and optimize its various cross-layer functions. In particular, AI techniques provide distributed, self-organizing solutions to complex wireless networking problems such as resource allocation, decoding/encoding, and cell association \cite{Deep_Oshea,Brendan_Federated}. Indeed, the association of users to BSs in ultra-dense and time-varying cellular networking environments becomes challengingf to model and solve mathematically while capturing all the network dynamics. This motivates the need for addressing cell association problems using distributed learning algorithms that enable both users and BSs to exploit the data that can be gathered by BSs in the network.

The privacy constraints in cellular networks coupled with the traffic load induced by centralized learning frameworks makes it necessary to develop distributed machine learning algorithms for cell association  \cite{Brendan_Federated}. These algorithms must be able to exploit the training data that is stored at a large number of devices to reduce the local training time and save their computing  and spectrum resources. The main objective when designing a distributed learning framework is to allow a given user to  benefit from the learning and processing of other neighboring users that have already selected their serving BS. For instance, users that are located in the same area will  often experience the same channel condition, and the same distance to the BS separates them. Thus, a given user can exploit information about selected BSs by users that have similar network conditions.

The problem of cell association was extensively addressed in the literature \cite{Elshaer_cell,Mozaffari_Optmal,Maghsudi_distributed} and \cite{Chen_Echo}. The work in \cite{Mozaffari_Optmal} addressed the problem of cell association between unmanned aerial vehicles (UAVs) and users using optimal transport theory to minimize the average network delay under any arbitrary spatial distribution of the ground users as well as the optimal cell partitions of UAVs and terrestrial base stations. The authors in \cite{Maghsudi_distributed} proposed a distributed cell association mechanism for energy harvesting IoT devices based on mean-field multi-armed games. In \cite{Chen_Echo}, the authors formulated the problem of cell association as a noncooperative game and proposed a distributed algorithm based on the machine learning framework of echo state networks (ESNs). The proposed algorithm enables the small base stations to autonomously choose their optimal bands allocation strategies while having only limited information on the states of the network and its users. The work in \cite{xx} also used machine learning to study cell association in cloud-based networks. Altough interesting, all these works either consider a static model or dynamic systems where all the information are assumed to be known to the BSs and users.

The main contribution of this paper is a novel collaborative learning  mechanism in ultra-dense cellular networks that can exploit the similarities between users in terms of network conditions. To this end, we introduce a new learning mechanism \emph{via imitation} that helps a user to select its serving BS faster by exploiting its local data and the learning outcomes of neighboring users. In fact, neighboring users might be characterized by similar characteristics such as channel conditions and their distance to the BSs. In this mechanism, instead of exchanging all the local data between users, only the outcome of their learning algorithms is transmitted.

In particular, we formulate the problem of cell association as a mean-field game (MFG) \cite{Gueant_MeanField} with imitation in which a user aims to maximize its own data rate while minimizing the cost of imitating its neighboring users. Then, we reduce the MFG into a Markov decision problem (MDP) which is essential for exploiting the measurements available at the users. Hence, learning which base stations the users should connect to as well as the reward function via local data becomes possible.

To reach the desirable mean-field equilibrium outcome for the formulated game, we propose a deep-learning based reinforcement learning algorithm that allows the users to predict their utility function by exploiting their local data and mimicking similar users. Using extensive simulations, we compare the proposed mechanism with a setting in which users select their serving BSs without imitating other similar users. Such a comparison allows us to see that these imitator users can learn the optimal action in a new environment faster than other users.

The rest of this paper is organized as follows. In Section \ref{sec:system_model}, we present the system model. In Section \ref{sec:problem_for}, the problem is formulated as a mean-field game with imitation and then a deep-learning based reinforcement learning algorithm is proposed to determine the user-cell association policy. Section \ref{sec:simul} presents the simulation results and Section \ref{sec:conclusion} concludes the work.  
\section{System Model}
\label{sec:system_model}
Consider a set $\mathcal{S}$ of $S$ small base stations (SBSs) deployed to serve a set $\mathcal{U}$ of $U$ users in an LTE cellular system. We consider both downlink and uplink of the LTE system. %(to 
We use $u \in \mathcal{U}$ and $s \in \mathcal{S}$ to index the users and SBSs, respectively. We introduce a binary variable $a_{su}$ that is equal to $1$ when user $u$ chooses SBS $s$ and $0$ when user $u$ is connected to another SBS. Each user $u$ decides to select SBS $s$ based on the following utility function:% (To Kenza: we can add another negative term for user power if we consider uplink. I was not sure about choosing which one of them) (I can use a weight for second term and decrease it over the time so when a user gains experience decrease its mimicking behavior)
\begin{equation}\label{eq:user_reward}
a_{su}=\text{arg}\max_s \left[ r_{us}-\sum_{u'=1}^U f(u,u')|a_{su}-a_{su'}|\right],
\end{equation}
where $r_{us}$ is the throughput of user $u$ when connected to SBS $s$, and $f$ is a function that captures the similarity between two users $u$ and $u'$. The similarity function captures the channel conditions, the geographical position and the interests of users. 

\begin{figure}[!t]
	\centering
	\includegraphics[scale=.6,trim={0.8cm 0 0 0}]{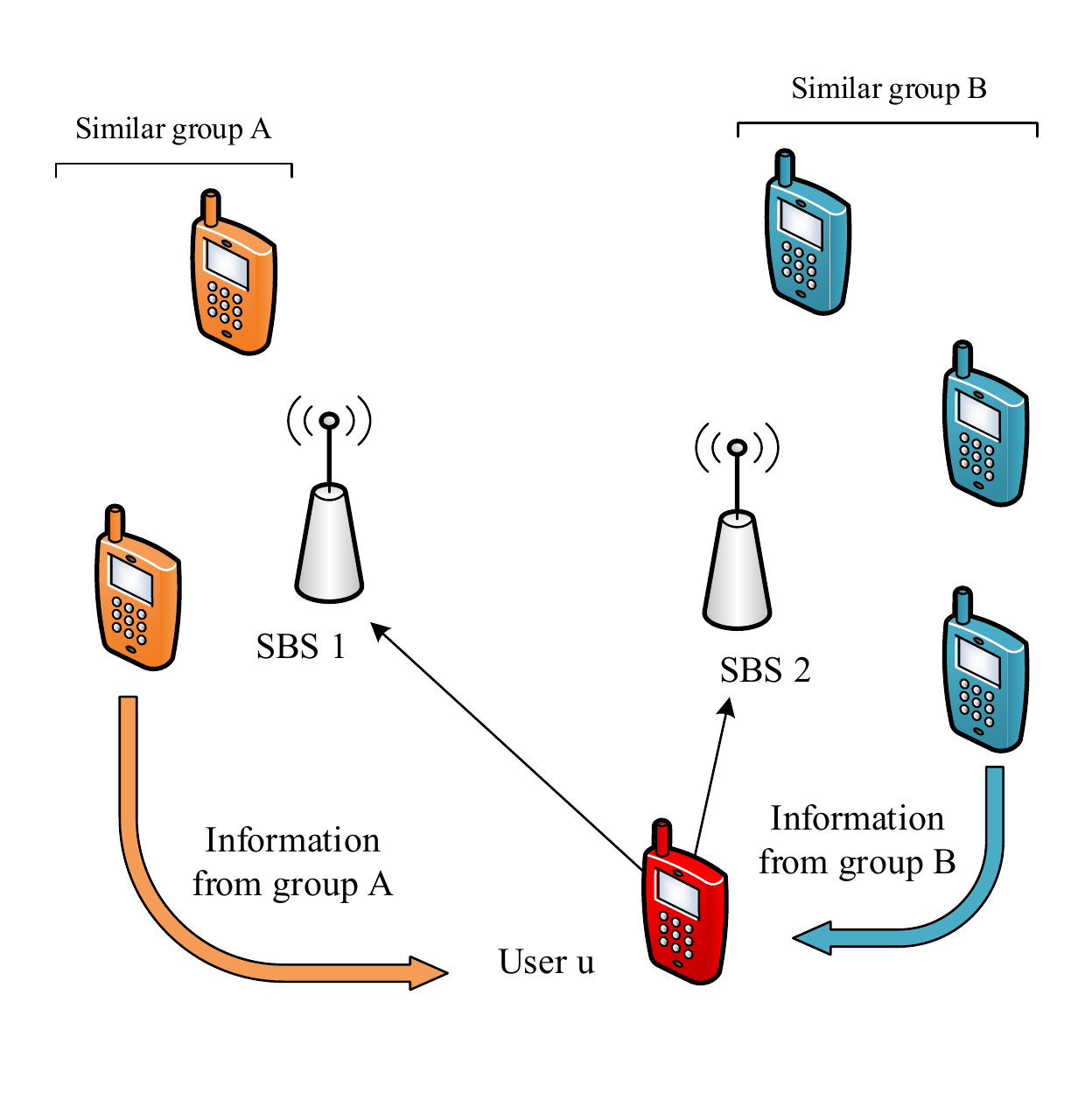}
	\caption{Illustration of our system model.}
	\label{fig:SysModel}
\end{figure}

The achievable throughput of user $u$ is given by 
\begin{equation}
r_{u,s}=B\log_2\left(1+\frac{p_{us} h_{us}}{\sigma^2+I(\sum_{u'} a_{su'})}\right)-c(\sum_{u'=1}^U a_{su'}),
\end{equation}
where $c(.)$ and $I(.)$ are increasing functions. $c(\sum_{u'=1}^U a_{su'})$  takes into account the throughput drop when the cell is congested. $I(\sum_{u'} a_{su'})$ determines interference in uplink and it is a function of the number of total users connected to the SBS $u$. $h_{us}$ is the channel gain between user $u$ and BS $s$. An illustration of the system model is shown in Fig. \ref{fig:SysModel}

In ultra-dense cellular systems, a large number of users deployed in the same area are covered by the same SBSs and experience the same quality-of-service (QoS) when the number of users tends to infinity. Thus, we introduce a similarity indicator function $f(u,u')$ and define it as follows:
\begin{equation}
f(u,u')=\begin{cases}
0, & \text{user $u$ is not similar to user $u'$, }\\
1, & \text{user $u$ is similar to user $u'$.}
\end{cases}
\end{equation}
The user distribution across SBSs at time slot $t$ is given by:
\begin{equation}
\boldsymbol \pi(t)=\begin{bmatrix}
\pi_1(t) &\cdots &\pi_S(t)
\end{bmatrix},
\end{equation}
where $\pi_s(t)$ is the fraction of users that are connected to SBS $j$ at time slot $t$. We can define transition probability
$ P_{sm}(t)$ as the probability that users connected to SBS $s$ switch to SBS $m$ in time slot $t$. Hence, the users' distribution evolves as follows: 
\begin{equation}
\pi_m(t+1)=\sum_{s=1}^S P_{sm}(t) \pi_s(t).
\end{equation}
The reward function for each user is defined as a function of $\boldsymbol \pi(t)$:
\begin{equation}\label{eq:reward_state}
r_u(\pi_s(t),s)= r_{us}(\pi_s(t))-\sum_{u'=1}^U f(u,u')|a_{su}-a_{Su'}|.
\end{equation}

Our primary goal is to assign the users to the SBSs while accounting for the high density of users in cellular systems and leveraging this density. To this end, we formulate the assignment problem as a mean-field problem in which the users exploit the storage and computing capabilities at the users to cooperatively decide to which SBSs they connect. 

\section{Problem Formulation and Game Analysis}
\label{sec:problem_for}
In this section, we formulate the problem of user-cell association as a mean-field game \cite{Gueant_MeanField,Brendan_Federated} with imitation to account for the high density of future cellular systems and leverage the data available at the users with a low communication overhead. Thus, we enable the users to collaborate and leverage both storage and processing capabilities that are locally available to them for an efficient cell association mechanism.

\subsection{Mean-Field Game Formulation}
Let $\mathcal{G}=(\mathcal{V},\mathcal{E})$ be a graph whose vertex set $\mathcal{V}$ represents the set of SBSs to which the users can connect and the edge set $\mathcal{E}$ represents the possible transition between every two SBSs. Thus, only the neighboring SBSs are connected with an undirected link. We define the state of  a given user $u\in \mathcal{U}$ at time $t$ as the SBS to which this user will get connected. 

For each SBS $s\in\mathcal{S}$, we define the two sets $\mathcal{V}_s=\{j:(j,i)\in\mathcal{E}\}$ and $\bar{\mathcal{V}}_s=\mathcal{V}_s\cup \{s\}$. The dynamics of the users are generated by right stochastic matrices $\boldsymbol{P}(t)\in S(\mathcal{G})$, where $\mathbb{S}(\mathcal{G})=\mathbb{S}_1(\mathcal{G})\times...\times \mathbb{S}_S(\mathcal{G})$ and each row $\boldsymbol{P}_s(t)$ belongs to $\mathbb{S}_s(\mathcal{G})=\{p\Delta^{S-1} |\textrm{supp}(p)\subset \bar{\mathcal{V}}_s\}$, where $\Delta^{S-1}$ is the simplex in $\mathbb{R}^S$. Moreover, we define a value function $V_s(t)$ of state $s$ at time $t$, and a reward function $r_s(\boldsymbol{\pi}(t),\boldsymbol{P}_s(t))$, quantifying the instantaneous reward of a user connected to SBS $s$ taking transitions with probability $P_s(t)$ when the current distribution of the users over the SBSs is $\boldsymbol{\pi}(t)$.

The backward Hamilton Jacobi-Bellman (HJB) equation and the forward Fokker-Planck equation for each SBS $s \in \{1, ..., S\}$ and time $t = 0,..., T-1$, in a discrete-time graph state MFG are given by:
\begin{equation}
V_s^t=\max_{\boldsymbol{P}_s^t\in S(\mathcal{G})}\left\{ r_s(\boldsymbol{\pi} (t),\boldsymbol{P}_s(t))+\sum_{j\in\bar{\mathcal{V}}_{s}}\boldsymbol{P}_{sj}(t) V_j(t+1)\right\},
\end{equation}
\begin{equation}
\boldsymbol{\pi}_s(t+1)=\sum_{j\in\bar{\mathcal{V}}} \boldsymbol{P}_{js}(t) \boldsymbol{\pi}_s(t).
\end{equation}
 
 Next, we define the elements that are necessary to formulate our problem.
\begin{itemize}
\item \emph{ Users distribution}  $\boldsymbol{p}_s(t)\in\Delta^{S-1}$ for $t=0,...,T-1$. Each $\boldsymbol{\pi}(t)$ is a discrete probability distribution over $S$ SBSs, where $\boldsymbol{\pi}_s(t)$ is the fraction of users that are connected to SBS $s$ at time $t$.
\item \emph{ Transition matrix} $\boldsymbol{P}(t)\in S(\mathcal{G})$. $\boldsymbol{P}_{sj}(t)$ is the probability that users connected to SBS $s$ switch to SBS $j$ at time $t$. We refer to $P_s(t)$ as the action of users conneced to SBS $s$. $\boldsymbol{P}(t)$ generated the forward equation
\begin{equation}
\pi_j(t+1)=\sum_{s=1}^S \boldsymbol{P}_{sj}(t)\pi_s(t).
\label{eq:forward}
\end{equation}
\item \emph{Reward} $r_s(\boldsymbol{\pi}(t),\boldsymbol{P}_s(t))=\sum_{j=1}^S \boldsymbol{P}_{sj}(t)r_{sj}(\boldsymbol{\pi}(t), \boldsymbol{P}_s(t))$ for $s\in\mathcal{S}$. This is the reward received by the users connected to SBS $s$ that choose action $\boldsymbol{P}_s(t)$ at time $t$, when the distribution is $\boldsymbol{\pi}(t)$. 
\item \emph{ Value function} $V(t)\in R^S$. $V_s(t)$ is the expected maximum total reward of being connected to SBS $s$ at time $t$. A terminal value $V^{T-1}$ is needed and will be set to zero.
\item \emph{ Average reward} $e_s(\boldsymbol{\pi},\boldsymbol{P},V)$, for $s\in\mathcal{S}$ and $V\in R^S$ and $P\in S(\mathcal{G})$. This is the average reward received by users connected to SBS $S$ when the current distribution is $\boldsymbol{\pi}$, action $\boldsymbol{P}$ is chosen, and the subsequent expected total reward is $V$. The average reward is defined as 
\begin{equation}
e_s(\boldsymbol{\pi},\boldsymbol{P},V)=\sum_{j=1}^S \boldsymbol{P}_{sj}(r_{sj}(\boldsymbol{\pi,\boldsymbol{P}})+V_j).
\end{equation}
\end{itemize}
Intuitvely, users want to act optimally in order to maximize their expected total average reward.

For $P\in S(\mathcal{G})$ and a vector $q\in S_s(\mathcal{G})$, we let $\mathcal{P}(\boldsymbol{P},s,q)$ be the matrix equal to $P$, where row $s$ is replaced by $q$. Then, we define the desirable outcome of the problem as follows.
\begin{definition}
A right stochastic matrix $P\in \mathbb{S}(\mathcal{G})$, is a Nash maximizer of $e(\boldsymbol{\pi},P,V)$, if given a fixed $\boldsymbol{\pi}$ and a fixed $V\in \mathbb{R}^S$, for any $s\in\mathcal{S}$ and any $q \in \mathbb{S}_s(\mathcal{G})$, there is 
\begin{equation}
e_s(\boldsymbol{\pi},\boldsymbol{P},V)\geq e_s(\boldsymbol{\pi},\mathcal{P}(\boldsymbol{P},s,q),V).
\end{equation}
\label{def:nash_max}
\end{definition}

The rows of $\boldsymbol{P}$ form a Nash equilibrium set of actions, since for any SBS $s$, the users connected to SBS $s$ cannot increase their reward by unilaterally switching their action from $\boldsymbol{P}_s$ to any $q$. Under Definition \ref{def:nash_max}, the value function of each SBS $s$ at each time $t$ satisfies the optimality criteria:
\begin{equation}
V_s(t)=\max_{q\in S_s(\mathcal{G})}\left( \sum_{j=1}^{S} q_j \left[r_{sj}(\boldsymbol{\pi}(t),\mathcal{P}(\boldsymbol{P}(t),s,q))+V_j(t+1)\right]\right).
\label{eq:optimality_cri}
\end{equation}
A solution of the MFG is a sequence of pairs $\{(\boldsymbol{\pi}(t),V(t))\}_{t=0,...,T}$ satisfying the optimality criteria (\ref{eq:optimality_cri}) and the forward equation (\ref{eq:forward}).
%%%%%%%%%%%%%%%%%%%%%%%%%%%%%%%%%%%%%%%%%

Now, we reduce the formulated MFG into a single user deterministic Markov decision process (MDP) within a finite time duration. This shows that solving the optimization problem of a single user MDP is equivalent to solving the MFG and allowing every user to select the SBS that maximizes its value function. This connection will allow us to apply efficient deep reinforcement learning (RL) methods that use data about the dynamics of the users in the network, to learn the best strategies of the users as well as their reward function.

\subsection{Mean-Field Game Analysis}
Here we formulate the problem as a Markov decision process for each user. Each user's action is defined as choosing an SBS to connect to. Its reward is defined in (\ref{eq:reward_state}). Also the state of the system $\boldsymbol x^t$ is defined as the number of users connected to each SBS:
\begin{equation}
\boldsymbol x(t) = \begin{bmatrix}
x_1(t) &\cdots &x_S(t)
\end{bmatrix},
\end{equation}
First, we need to find number of states for the MDP as follows.
\begin{proposition}\label{prop:size_ss}
Let the number of SBSs that each user $u$ can use be $b_u\leq S$. Also, let the total number of users that can connect to SBS $s$ be $N_s$. The total number of states for user $u$ is $S_u$ and is bounded as follows:
\begin{equation}
K_u \leq \binom{\sum_{s=1}^{b_u} N_s-b_u+1}{b_u-1} \leq \binom{U-b_u+1}{b_u-1}
\end{equation}
\end{proposition}
\begin{IEEEproof}
The total number of states for a given user $u$ is given by the non-negative integral solutions of the following equation:
\begin{equation}
K_u=\sum_{s=1}^{b_u} n_s= \sum_{s=1}^{b_u} N_s,
\end{equation}
where $n_s$ is actual number of users connected to SBS $s$. Hence, we can find the upper limit for $K_u$ as, 
\begin{equation}
K_u \leq \binom{\sum_{s=1}^{b_u} N_s-b_u+1}{b_u-1}.
\end{equation}
Furthermore, since 
\begin{equation}
 \sum_{s=1}^{b_u} N_s\leq U,
\end{equation}
we know that
\begin{equation}
\binom{\sum_{s=1}^{b_u} N_s-b_u+1}{b_u-1} \leq \binom{U-b_u+1}{b_u-1}.
\end{equation}
\end{IEEEproof}

As we can see from Proposition \ref{prop:size_ss}, the size of the state space grows with the number of users in the order of $\mathcal{O}(U^{b_u})$ in the worst case. Since each agent uses Q-learning for learning the optimal action, it has to store the Q-function. However, as we can see from Proposition \ref{prop:size_ss}, it is not feasible to create a table for the Q-function. 
%We use a function to represent value function $V$. 
%With the assumption of full observability for the users the problem is trivial to solve. However, we do not want to make such strong assumption. It is impossible for each user to find out number of users connected to other base stations. 
The only assumption we make on the system is that each user knows its reward after connecting to each SBS. A user can estimate $\sum_{u'=1}^U a_{su'}$ based only on its own reward.
We know 
\begin{align}
&\frac{\partial r_{u,s}}{\partial \sum_{u'=1}^K a_{su'}}=-c'(\sum_{u'=1}^U a_{su'}) \nonumber\\
&-\frac{p_{ij}h_{ij} I'(\sum_{u'=1}^K a_{su'})}{\big(\sigma^2+I(\sum_{u'=1}^K a_{su'})\big) \big( \sigma^2+I(\sum_{u'=1}^U a_{su'})+p_{ij}h_{ij}\big)},
\end{align}
and we know that $\frac{d I(x)}{x}>0$ and $\frac{d c(x)}{x}>0$. Hence, using the bisection method with the knowledge of $r_{u,s}$ each user can find total number of users connected to the BS.

Since each user cannot observe the full state of the system and it only observes it partially, we propose a method to solve partially observable Markov decision processes (POMDPs). In this method, each user estimates the full state using its limited observations and  a neural Q-network. We use multilayer neural networks as Q-function estimator.
  
At each time slot $t$, all the users make decision using reinforcement learning and estimate their Q-function using multilayer neural network.

\subsection{Value function approximation}
As we showed, the value function cannot be stored in a table. Therefore, a function approximation method should be used to approximate the value function. Neural networks are powerful tools for value function approximation \cite{mnih-dqn-2015}.  

Since each user does not know the transition model of the MDP, they need to approximate the Q-function instead of the value function. Considering the fact that training adaptive linear neurons using the backpropagation algorithm is computationally inexpensive, we can use a unique  neural network $n$ for approximating $Q(e,a_n)$. $e$ is the partial state observed by the user in the system and $a_n$ is the action, i.e., the SBS selected by the user. In this neural network, the state of the system is the input to the neural network. This is due to the fact that the number of actions for each user is limited in contrast to the large number of states.

The output of each adaptive linear neuron can be written as 
\begin{equation}
Q_u(e,a_n)=\boldsymbol w_n^T \boldsymbol x+\boldsymbol b_n.
\end{equation}
The approximating process is to choose a random  action at each stage and then trying to update the weights. To do so, we update the weights based on the following rule:
\begin{equation}
\boldsymbol w_n(t+1)=\boldsymbol w_n(t)+\lambda (y_u-Q_u(e,a_n)),
\end{equation}
where $\lambda$ is the learning rate, and $y$ is the current target which is an exponential moving average and can be written as 
\begin{equation}
y_u=(1-\alpha)Q_u(e,a_n)+ \alpha (r(e,a_n)+ \gamma \max_{n}\boldsymbol w_n^T \boldsymbol x+\boldsymbol b_n),
\end{equation}
where $\alpha$ is a factor between 0 and 1. We use these adaptive linear neurons in a multilayer structure to estimate the Q-function, using which we find the optimal action as follows:
\begin{equation}\label{eq:best_bs}
a_{u}(e)=\max_{n} Q_u(e,a_n).
\end{equation}
That is, each user $u$ finds its best action $a_{u}(e)$ in state $e$ using the maximum output of its neural networks. Each neural network estimates the value of an action in the current partially observed state.

\section{Simulation Results and Analysis}\label{sec:simul}

  \begin{figure}[!t]
	\centering
		\includegraphics[scale=.7,trim={0.8cm 0 0 0}]{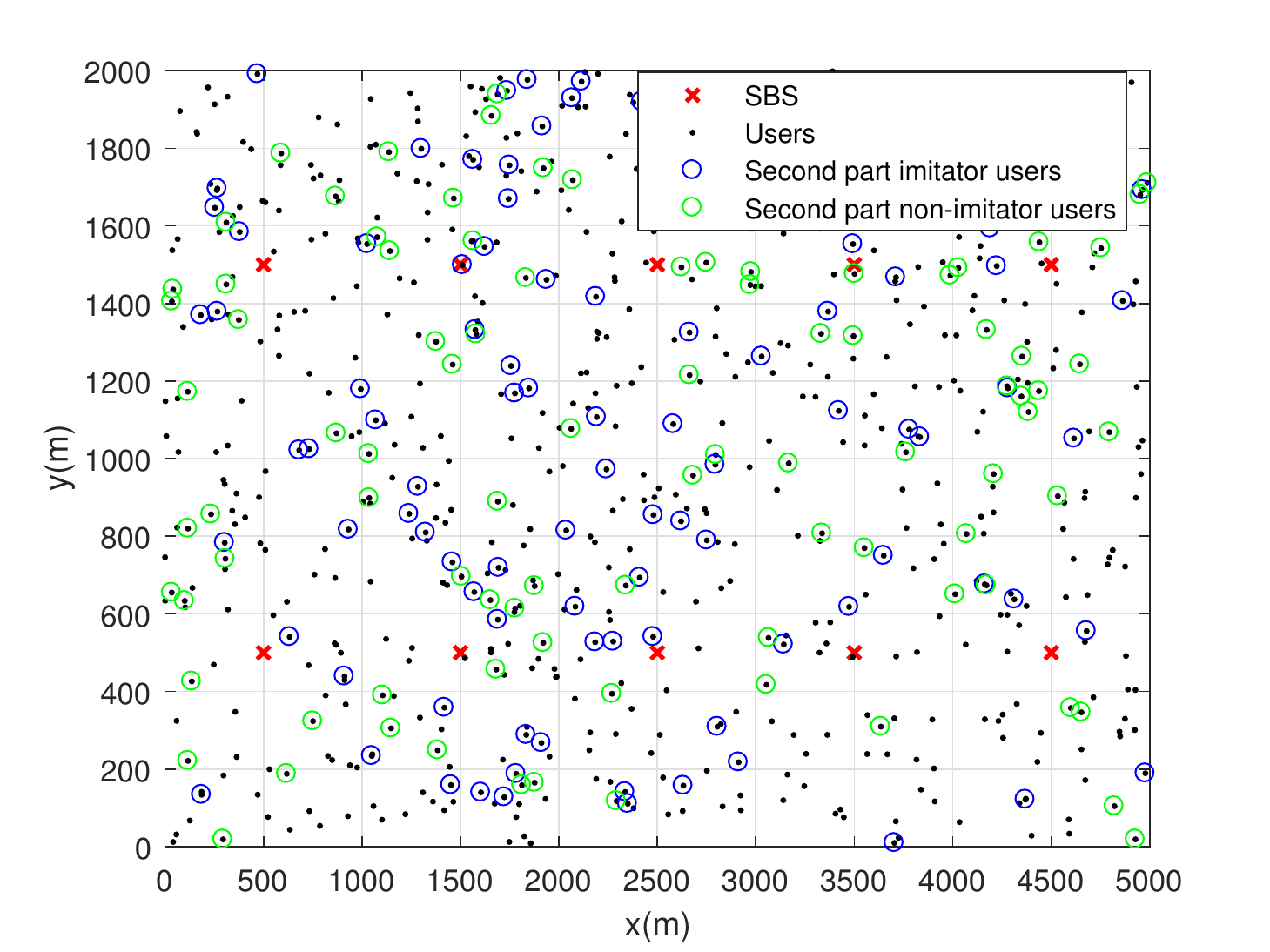}
	\caption{ٍWireless network with 10 SBS and 700 users.}
	\label{fig:location}
\end{figure} 

	For our simulation, we consider a network with 700 users  uniformly distributed in the range of 10 SBSs.
 
 We assume that the path loss exponent is 2, the carrier frequency is $900$~MHz, and the noise variance is $-173.9$~dbm/Hz. Each user acts based on an $\epsilon$-greedy reinforcement learning, meaning that it chooses a random action with probability $\epsilon$ and approximates the value function using adaptive linear neuron (ADALINE) neural networks. The user, then, trains its network using the Widrow-Hoff algorithm (exploration) and  chooses the best BS with probability $1-\epsilon$ using [\ref{eq:best_bs}] (exploitation).  
 \begin{figure}[!t]
	\centering
	\includegraphics[scale=.7,trim={0.6cm 0 0 0}]{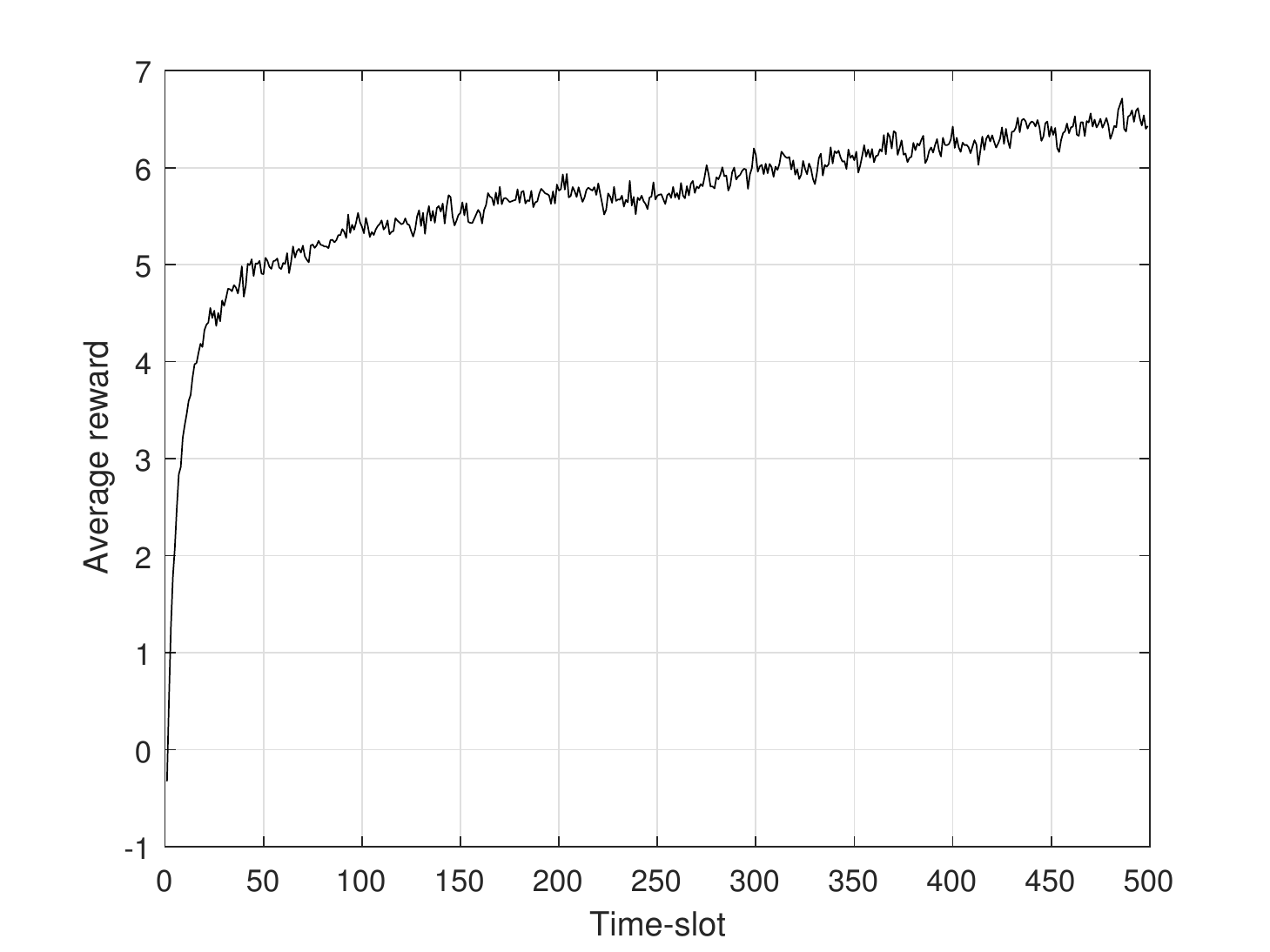}
	\caption{ٍAverage reward for 500 users in the first phase.}
	\label{fig:first_part}
\end{figure}    
 
\begin{figure}[!t]
	\centering
	\includegraphics[scale=.6,trim={0.2cm 0 0 0}]{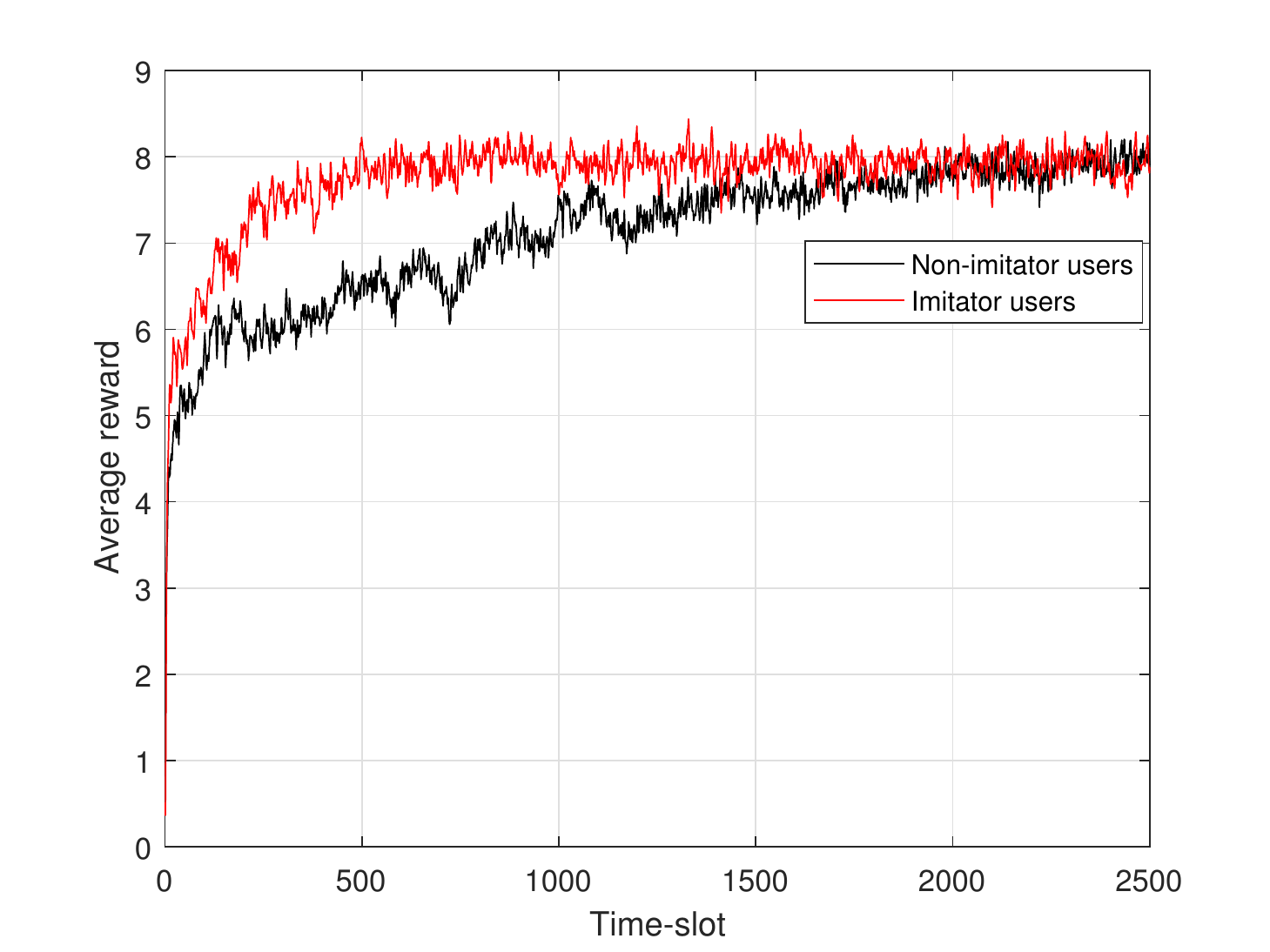}
	\caption{ٍTransient time of average reward for 100 imitator and 100 non-imitator users in the second phase.}
	\label{fig:second_part}
\end{figure}

 There are two different types of users in the learning algorithm:
 \begin{enumerate}
 \item \emph{Non-imitator users}: users that maximize a reward function affected only by the throughput and congestion functions. 
 \item \emph{Imitator users}: users that maximize the reward function of  non-imitator users in addition to imitating the action of the users close to them. 
 \end{enumerate}
 The simulation consists of two phases. In the first phase, 500 non-imitator users start to approximate and  maximize their value function using the aforementioned methods. After they converge to an equilibrium, in the second phase, 100 non-imitator and 100 imitator users enter the system. As we can see in Fig. \ref{fig:first_part}, non-imitator users start to learn the environment in $500$~time-slots and their average reward increases with time. This is due to the fact that, as they learn to coordinate with each other and manage interference, their average reward will increase.  Fig. \ref{fig:second_part} shows the second phase of the simulation where $100$ imitator users and $100$ non-imitator users enter the system and start to approximate a value function and maximize it.  The locations of the users, imitator users, non-imitator users and SBSs in the system are depicted in Fig. \ref{fig:location}.
 
\begin{figure}[!t]
	\centering
	\includegraphics[scale=.6,trim={0.8cm 0 0 0}]{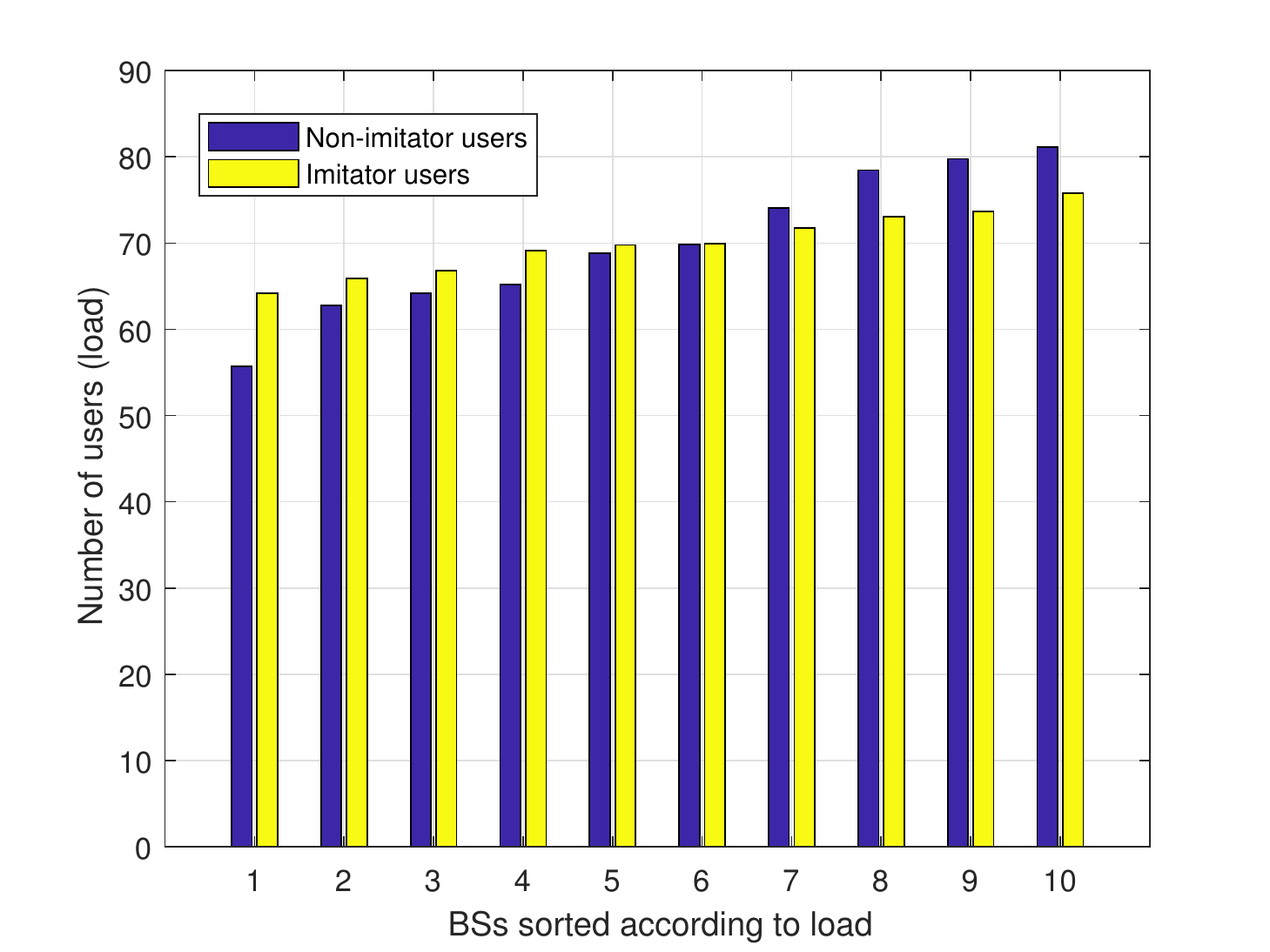}
	\caption{ٍLoad balance for each SBS in case of imitator and non-imitator users.}
	\label{fig:barplot}
\end{figure}

 As we can see, the imitator users adapt faster to the environment. This is due to the fact that, in addition to learning the environment, the imitator users   also use the existing users' experience. This experience will help them to learn the environment and behavior of other users faster. Since non-imitator users will also gain experience over time, the average reward of non-imitator users will  eventually approach that of imitator users. However, non-imitator users gain the experience of the existing users with a delay which is in a  direct relationship with the experience of their adjacent users. 
The average number of users per base station in two different cases is shown  in Fig. \ref{fig:barplot}. After 500 iterations, in case 1, 200 imitator users enter the system, and in case 2, 200 non-imitator users enter the system. Then, we find  the average number of users per base station for 100 time-slots. Since the imitator users use the existing knowledge of 500 users in the system, they can learn to adapt  with the system faster, and as a result, the load is more evenly balanced in this case.

\section{Conclusion}
\label{sec:conclusion}
In this paper, we have addressed the problem of cell association in ultra-dense networks while leveraging the data available at the users and their processing power. First, we have formulated the problem as a mean-field game with imitation, where users not only learn from their own local data, but also from the models learned by their neighboring users with the same characteristics, captured by a similarity function. We have showed via simulations that the proposed collaborative learning mechanism outperforms the learning mechanism without imitation in terms of learning time and load balance.

\end{document}